\documentclass[onecollarge,natbib]{svjour2}
\bibpunct{[}{]}{;}{n}{}{,} 
\smartqed  
\usepackage{graphicx}

\newcommand{\be}{\begin{eqnarray}}
\newcommand{\ee}{\end{eqnarray}}
\newcommand{\eq}{\begin{eqnarray}}
\newcommand{\en}{\end{eqnarray}}

\newcommand{\bfq}{{\bf q}_{\perp}}

\newcommand{\bfk}{{\bf k}_{\perp}}

\journalname{Few-Body Systems}
\begin{document}

\title{A Comparative Study of Nucleon Structure in Light-Front Quark Models in AdS/QCD}

\author{Chandan Mondal         \and
        Dipankar Chakrabarti 
}


\institute{Chandan Mondal and Dipankar Chakrabarti \at
Department of Physics, Indian Institute of Technology Kanpur,
              Kanpur 208016, India\\
              \email{mchandan@iitk.ac.in}, \email{dipankar@iitk.ac.in}            
}

\date{Received: date / Accepted: date}

\maketitle

\begin{abstract}
We present a comparative study of nucleon structure such as electromagnetic form factors, transverse charge and magnetization densities  in three different models within AdS/QCD framework.
\end{abstract}

\section{Introduction}\label{intro}
In recent years, AdS/QCD has emerged as one of the most promising techniques to unravel the structure of mesons and nucleons. AdS/CFT conjecture relates a strongly coupled gauge theory in $d$ space-time dimensions by a dual weak coupling gravity theory in AdS$_{d+1}$ space. 
To exploit this duality to address the  problems in QCD, the conformal invariance needs to be broken. In the literature there are two methods to achieve this goal, one is called hard wall model where a sharp cut-off is put in the hologhaphic direction in the AdS space where the wave functions are made to vanish and the other is called the soft wall model in which a confining potential is introduced in the AdS space  which breaks the conformal invariance and allows the QCD mass scale.

Electromagnetic form factors provide us insights into the structure of the nucleons and have been measured in many experiments. 
In the light-cone frame with $q^+=q^0+q^3=0$, the charge and anomalous magnetization densities in the transverse plane can be identified with the two-dimensional Fourier transform(FT) of the electromagnetic form factors. The contributions of individual quark to the nucleon charge and magnetization densities are obtained from the flavor decompositions of the transverse densities.

Here we present a detailed analysis of the nucleon form factors   in AdS/QCD soft-wall models \cite{AC,BT2,CM2,chandan}. The flavor form factors are obtained by decomposing the Dirac and Pauli form factors for nucleons using the charge and isospin symmetry.
 We also present a comparative study of the nucleon as well as the flavor contributions to the nucleon charge and anomalous magnetization densities in the transverse plane~\cite{CM3}. 
\section{Nucleon and flavor form factors in ADS/QCD models} \label{sec:1}
{\bf Model I} : 
Model I refers to the AdS/QCD model for nucleon form factors proposed by Brodsky and T\'{e}ramond \cite{BT2}.
The relevant AdS/QCD action for the fermion field is written as 
\be
S&=&\int d^4x dz \sqrt{g}\Big( \frac{i}{2}\bar\Psi e^M_A\Gamma^AD_M\Psi -\frac{i}{2}(D_M\bar{\Psi})e^M_A\Gamma^A\Psi-\mu\bar{\Psi}\Psi-V(z)\bar{\Psi}\Psi\Big),\label{action}
\ee
where $e^M_A=(z/R)\delta^M_A$ is the inverse  vielbein and $V(z)$ is the confining potential which breaks the conformal invariance and 
 $R$ is the AdS radius. 
In $d=4$ dimensions, $\Gamma_A=\{\gamma_\mu, -i\gamma_5\}$.
To map the Dirac equation in AdS space with the light front wave equation, one identifies $z\to\zeta$ (light front transverse impact variable) and substitutes $\Psi(x,\zeta)=e^{-iP\cdot x}\zeta^2\psi(\zeta)u(P)$  and sets $\mid \mu R\mid=\nu+1/2$ where  $\nu=L+1$ (more details can be found in \cite{BT2}).
For linear confining potential  $U(\zeta)=(R/\zeta)V(\zeta)=\kappa^2\zeta$, one gets the light front wave equation for the baryon which leads to the AdS solutions of nucleon wavefunctions $\psi_+(z)$ and $\psi_-(z)$ corresponding to different orbital angular momentum $L^z=0$ and $L^z=+1$ \cite{BT2}
\be
\psi_+(\zeta)\sim\psi_+(z) = \frac{\sqrt{2}\kappa^2}{R^2}z^{7/2} e^{-\kappa^2 z^2/2}\label{psi+},\quad\quad
\psi_-(\zeta)\sim\psi_-(z) = \frac{\kappa^3}{R^2}z^{9/2} e^{-\kappa^2 z^2/2}\label{psi-}.
\ee 
The Dirac form factors in this model are obtained by the SU(6) spin-flavor symmetry and given by 
\be
F_1^p(Q^2)=R^4\int \frac{dz}{z^4} V(Q^2,z)\psi^2_+(z),\label{F1p} \quad 
F_1^n(Q^2) = -\frac{1}{3}R^4\int \frac{dz}{z^4} V(q^2,z)(\psi^2_+(z)-\psi^2_-(z)).\label{F1n}
\ee
A precise mapping for the spin-flip nucleon form factor using the action in Eq.(\ref{action}) is not possible. Thus, the Pauli form factors for the nucleons are modeled in this model as   
\be
F_2^{p/n}(Q^2) =  \kappa_{p/n}R^4 \int \frac{dz}{z^3}\psi_+(z) V(Q^2,z) \psi_-(z).\label{F2}
\ee
The Pauli form factors are normalized to $F_2^{p/n}(0) = \kappa_{p/n}$ where $\kappa_{p/n}$ are the anomalous magnetic moment of proton/neutron.  $V(Q^2,z)$ is the bulk-to-boundary propagator\cite{BT2}.
Here we use the value $\kappa=0.4 GeV$ which is fixed by fitting the ratios of Pauli and Dirac form factors for proton with the experimental data \cite{CM2,CM1}. 
\begin{figure*}[htbp]
\begin{minipage}[c]{0.98\textwidth}
\small{(a)}
\includegraphics[width=5.5cm,clip]{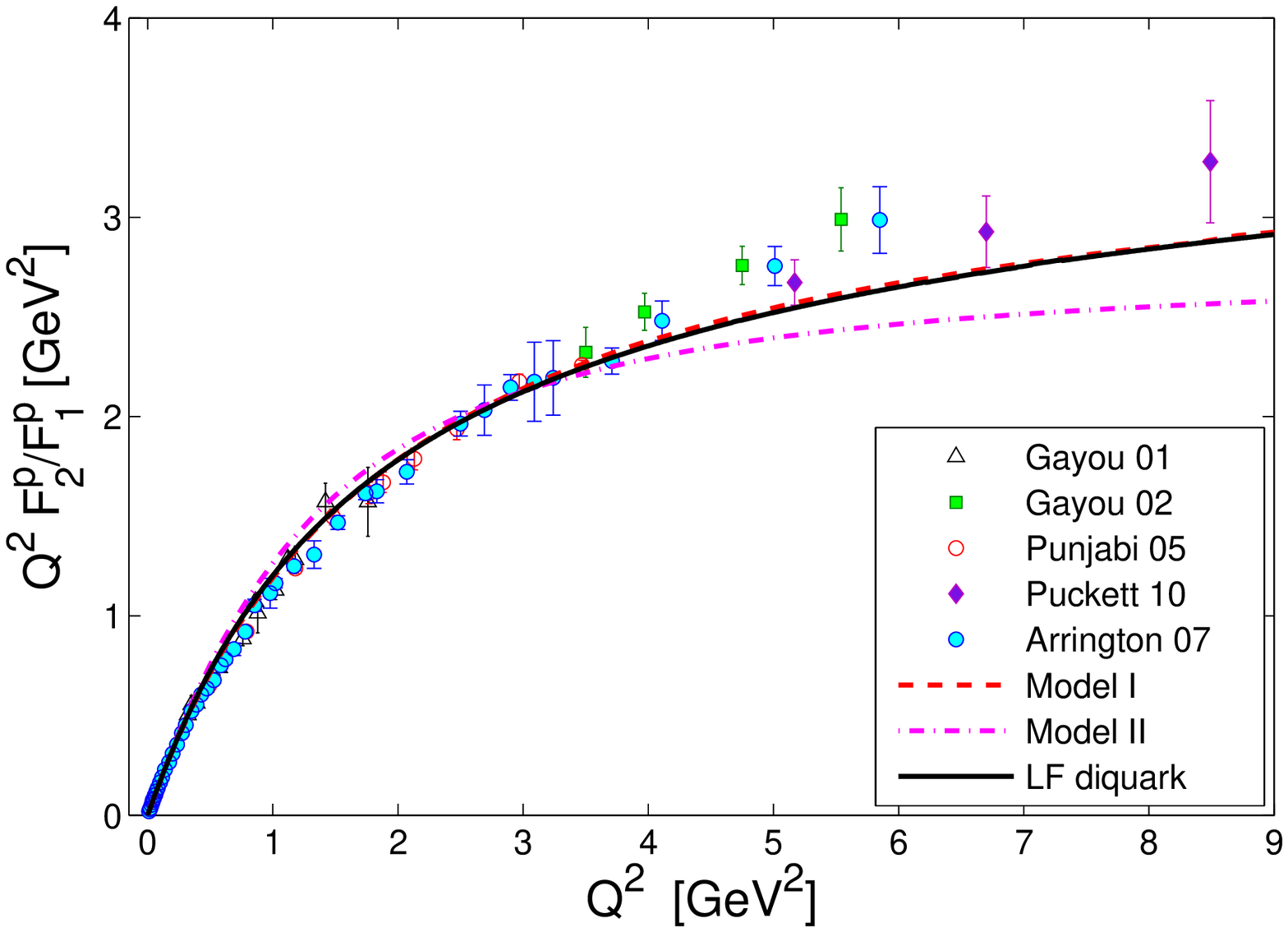}
\hspace{0.1cm}%
\small{(b)}\includegraphics[width=5.5cm,clip]{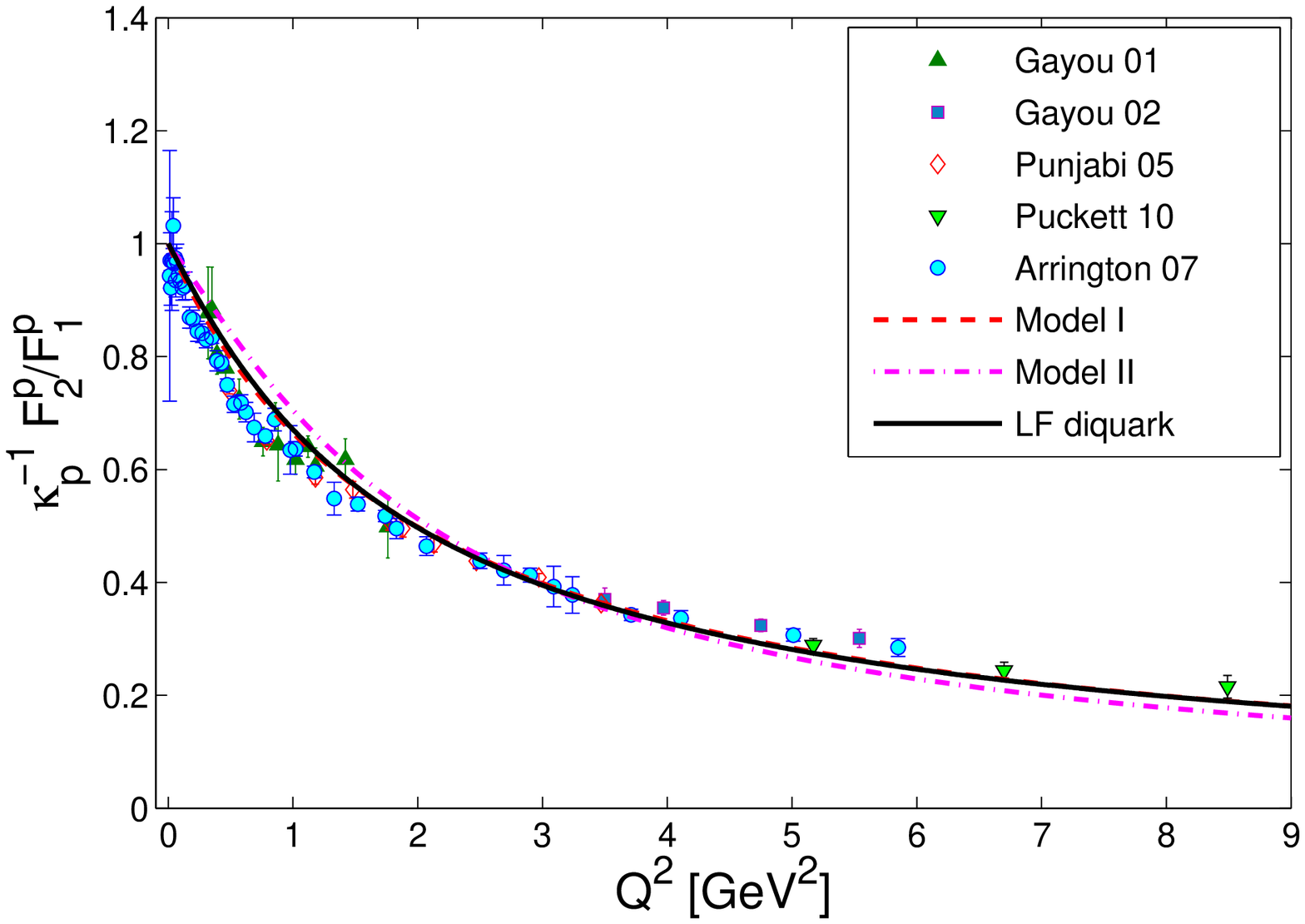}
\end{minipage}
\begin{minipage}[c]{0.98\textwidth}
\small{(a)}
\includegraphics[width=5.5cm,clip]{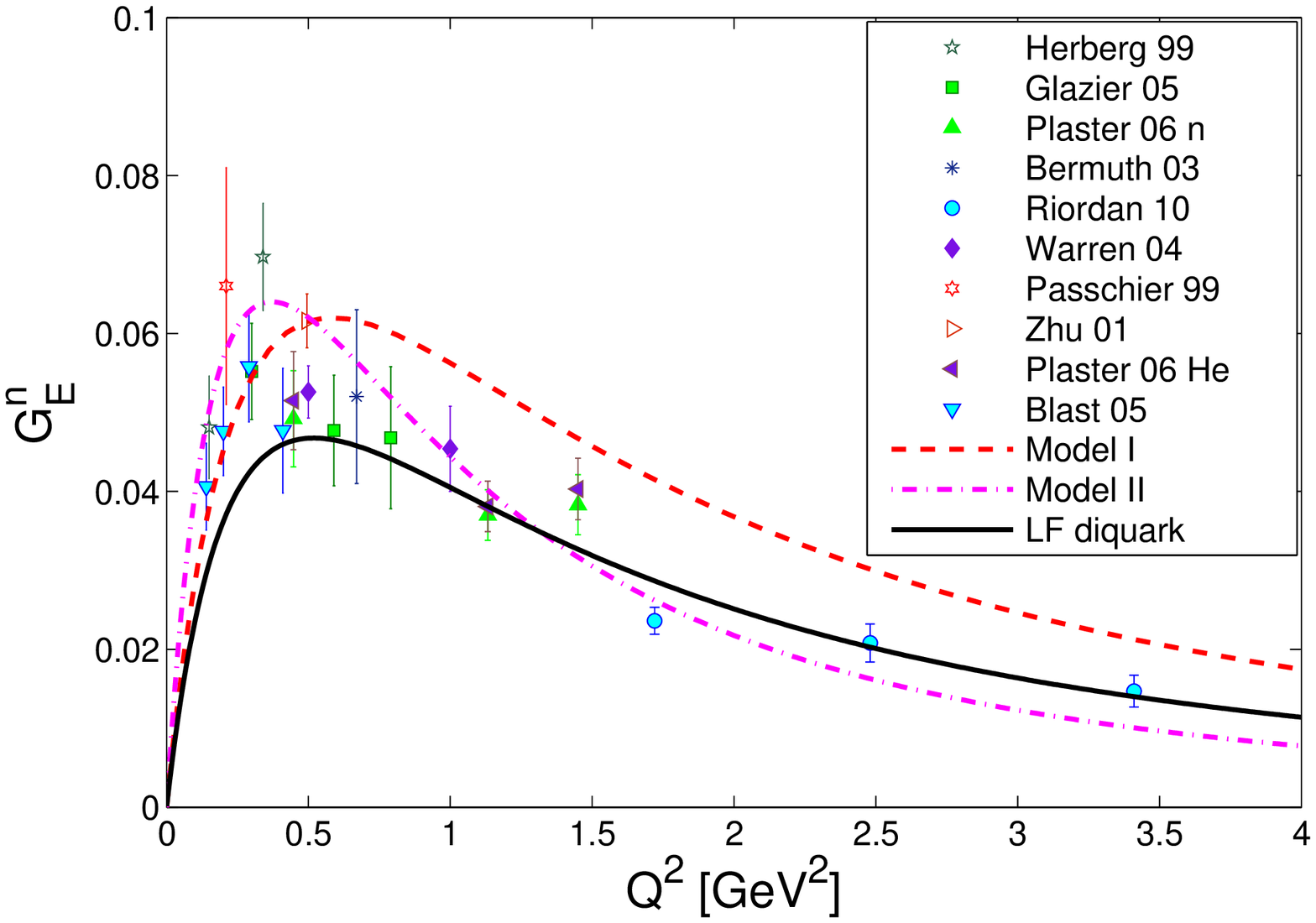}
\hspace{0.1cm}%
\small{(b)}\includegraphics[width=5.5cm,clip]{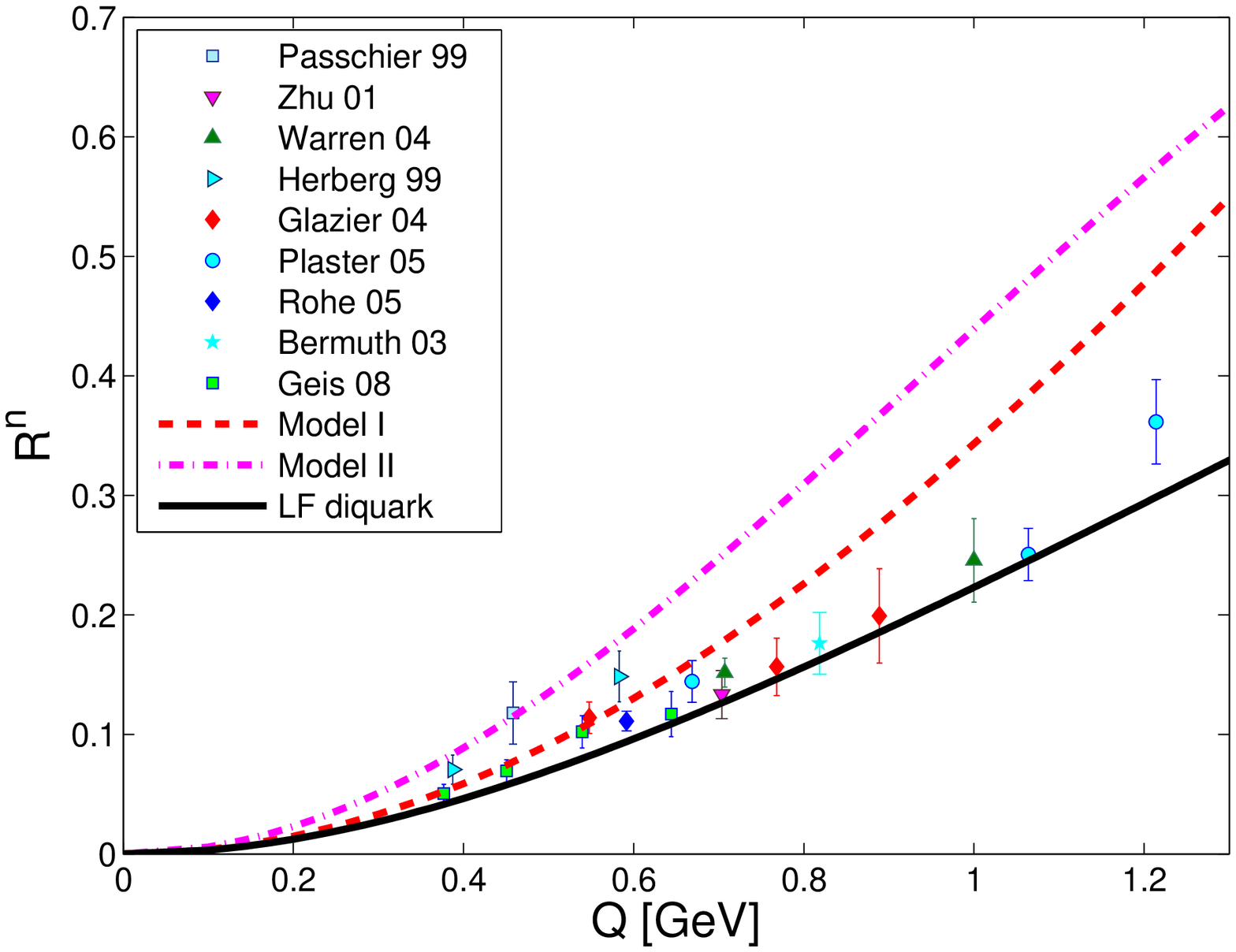}
\end{minipage}
\caption{\label{proton_fit}
Nucleon form factors (a) the ratio of Dirac and Pauli form factors for proton is multiplied by $Q^2=-q^2=-t$, (b) the ratio is divided by $\kappa_p$; (c) electric Sach form factor for neutron $G_E^n(Q^2)$, (d) the ratio $R^n=\frac{\mu_n~G_E^n}{G_M^n}$. 
The red dashed and pink dash dot lines represent the Model I and Model II and the solid black lines represent the quark-diquark model \cite{chandan}.
The references of the experimental data can be found in Ref.\cite{CM2}.}
\end{figure*}
\begin{figure*}[htbp]
\begin{minipage}[c]{0.98\textwidth}
\small{(a)}
\includegraphics[width=5cm,clip]{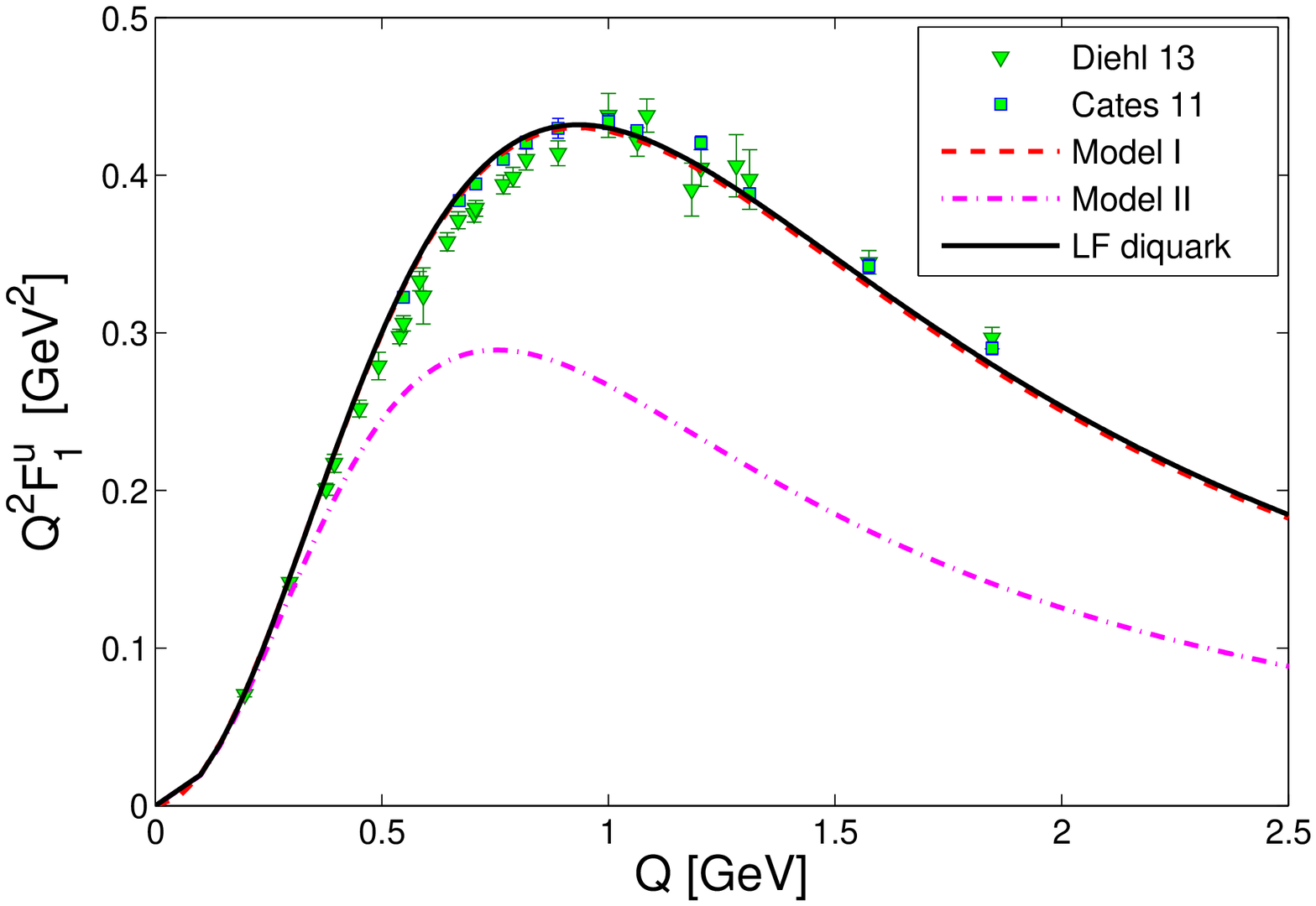}
\hspace{0.1cm}%
\small{(b)}\includegraphics[width=5cm,clip]{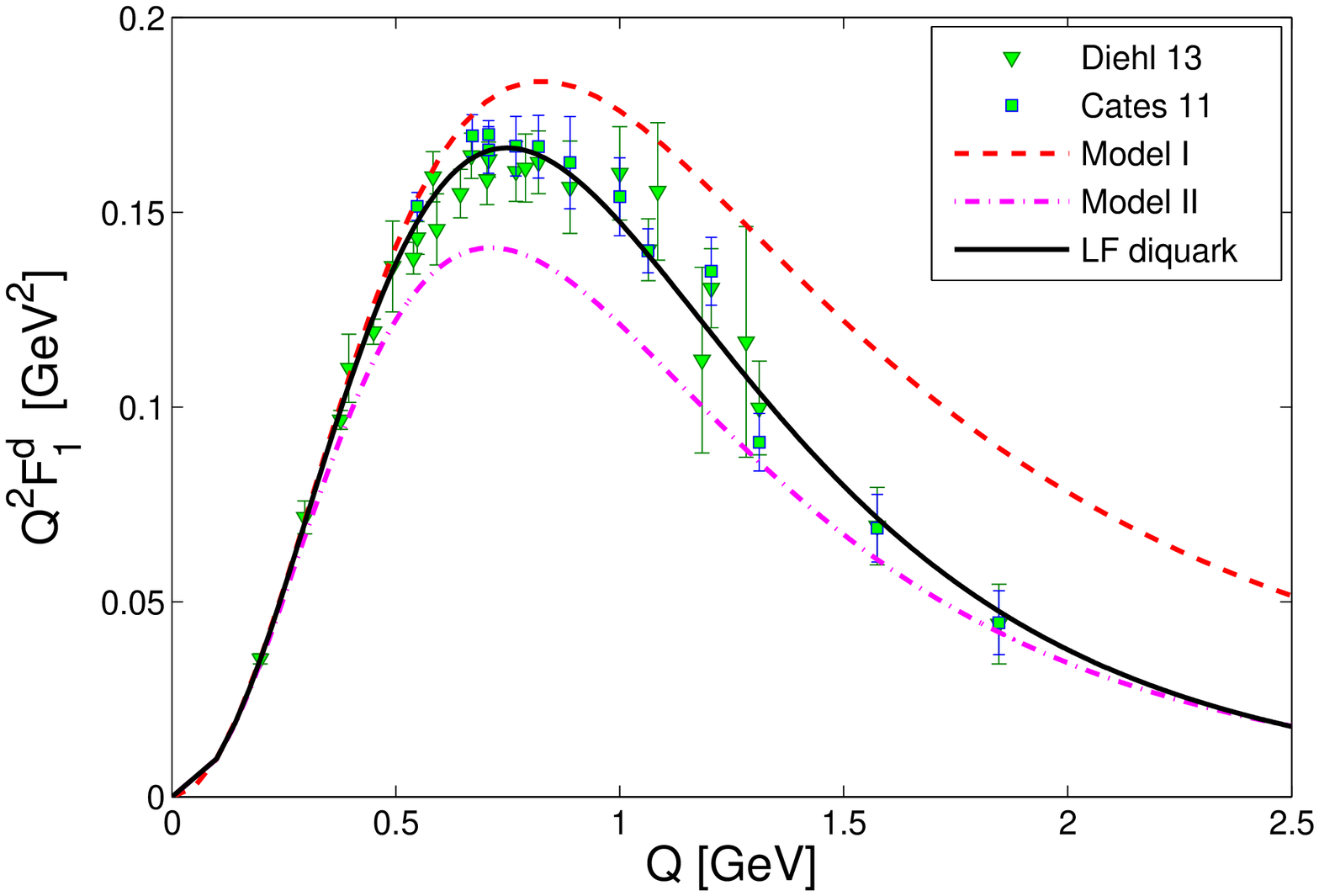}
\end{minipage}
\begin{minipage}[c]{0.98\textwidth}
\small{(c)}
\includegraphics[width=5cm,clip]{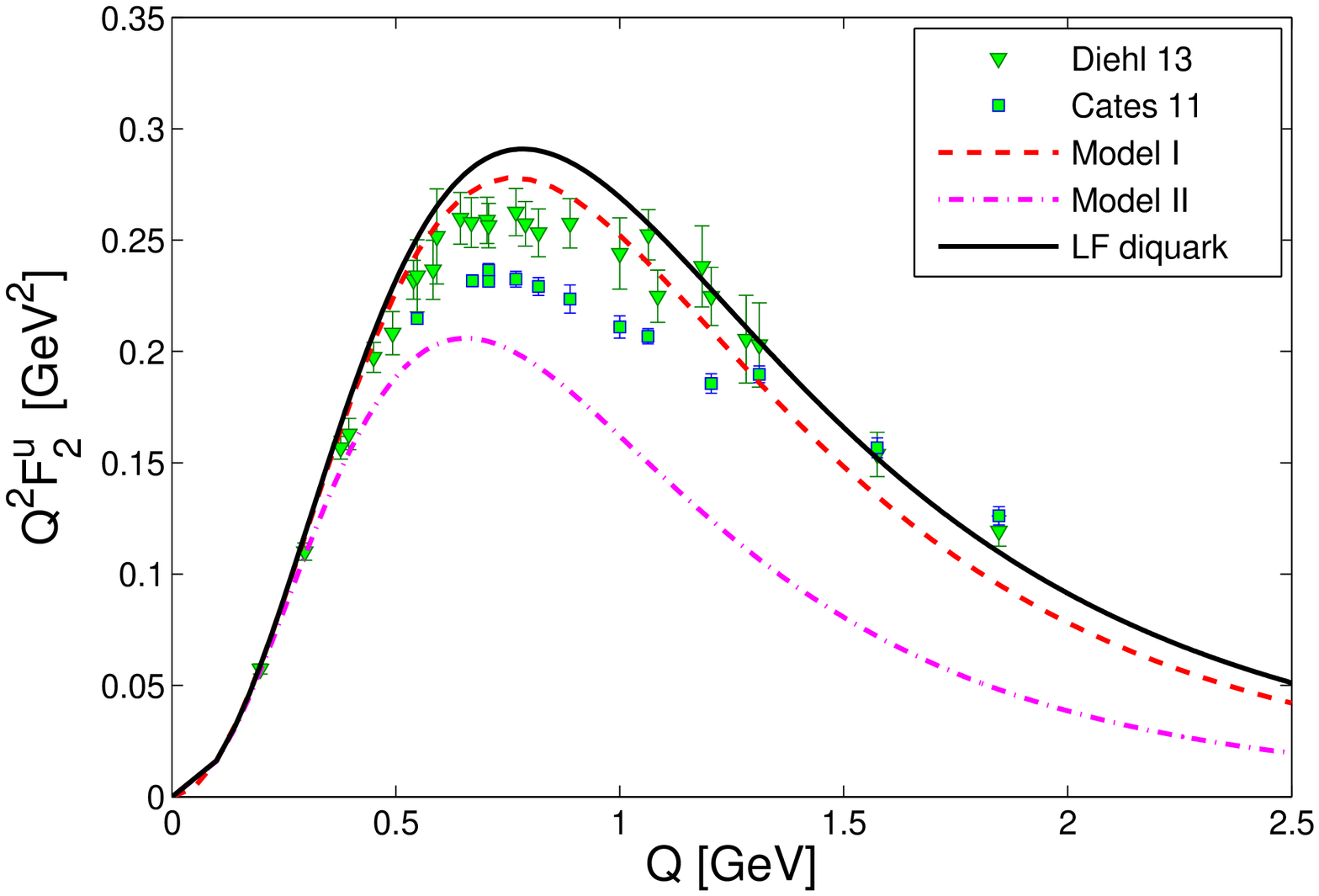}
\hspace{0.1cm}%
\small{(d)}\includegraphics[width=5cm,clip]{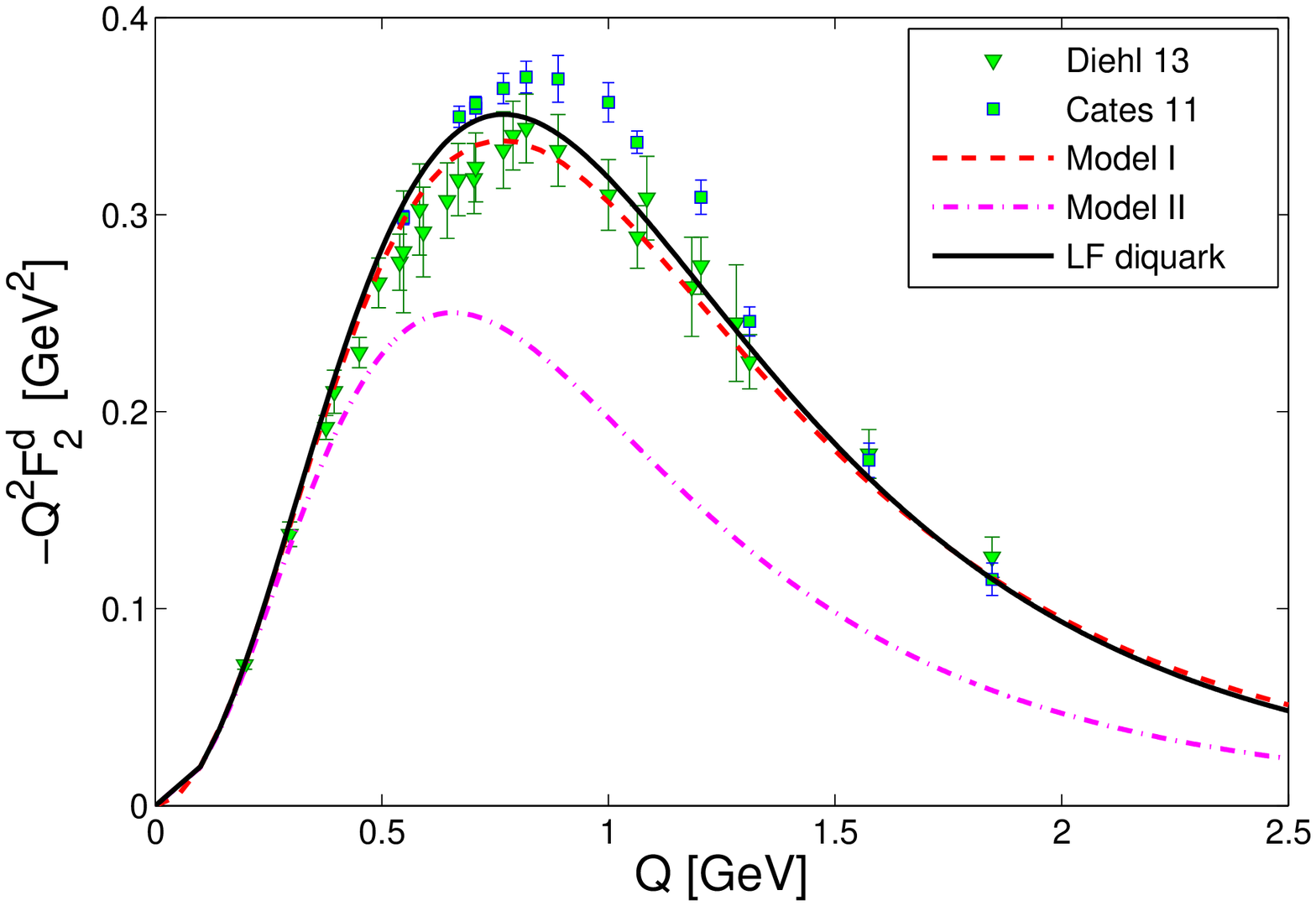}
\end{minipage}
\caption{\label{FF_flavors} Flavor dependent form factors for $u$ and $d$ quarks. The experimental data are taken from \cite{Cates,diehl13}. The lines are same as in Fig.\ref{proton_fit}.}
\end{figure*} 
\begin{figure*}[htbp]
\begin{minipage}[c]{0.98\textwidth}
\small{(a)}
\includegraphics[width=5cm,clip]{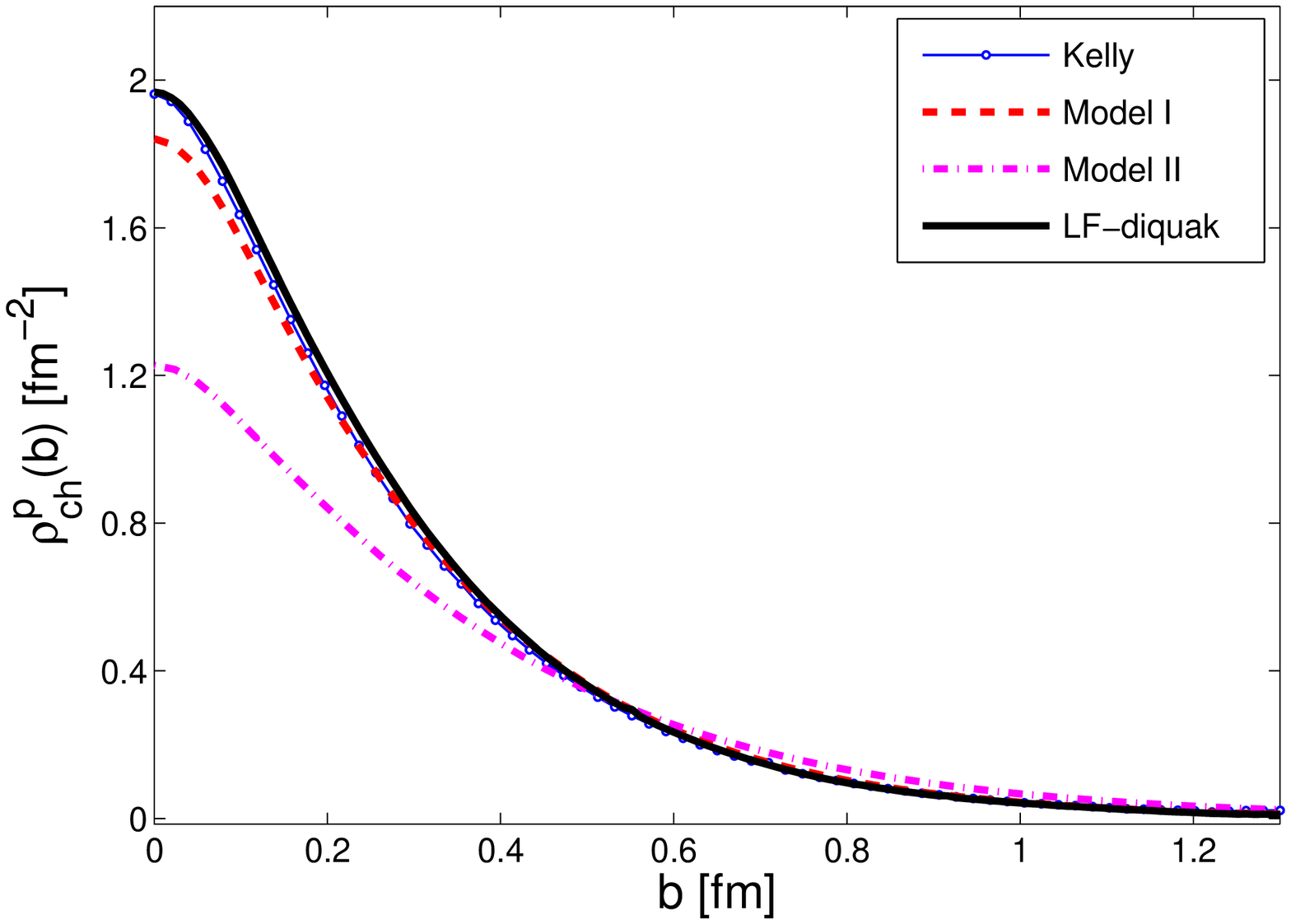}
\hspace{0.1cm}%
\small{(b)}\includegraphics[width=5cm,clip]{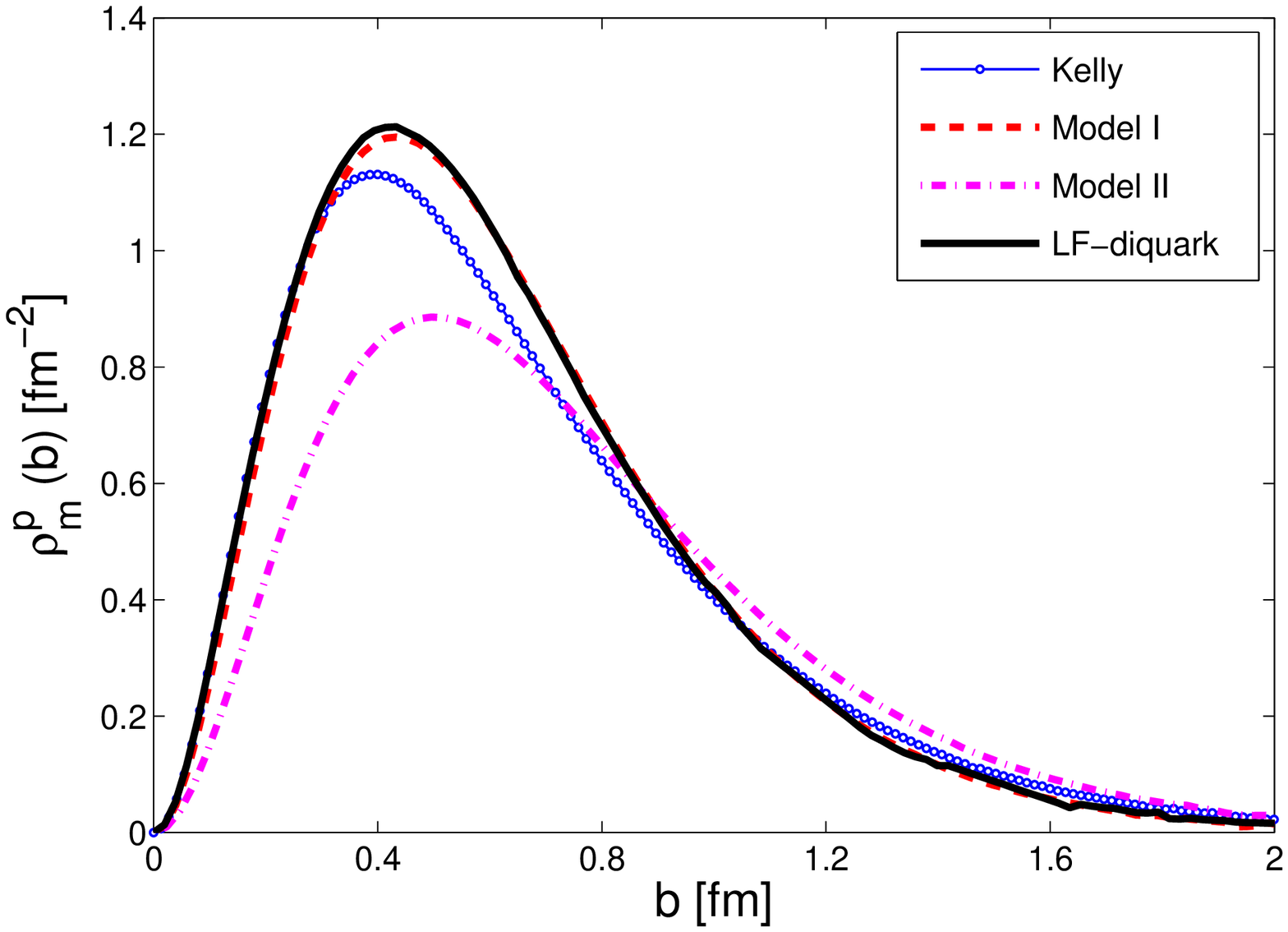}
\end{minipage}
\begin{minipage}[c]{0.98\textwidth}
\small{(c)}
\includegraphics[width=5cm,clip]{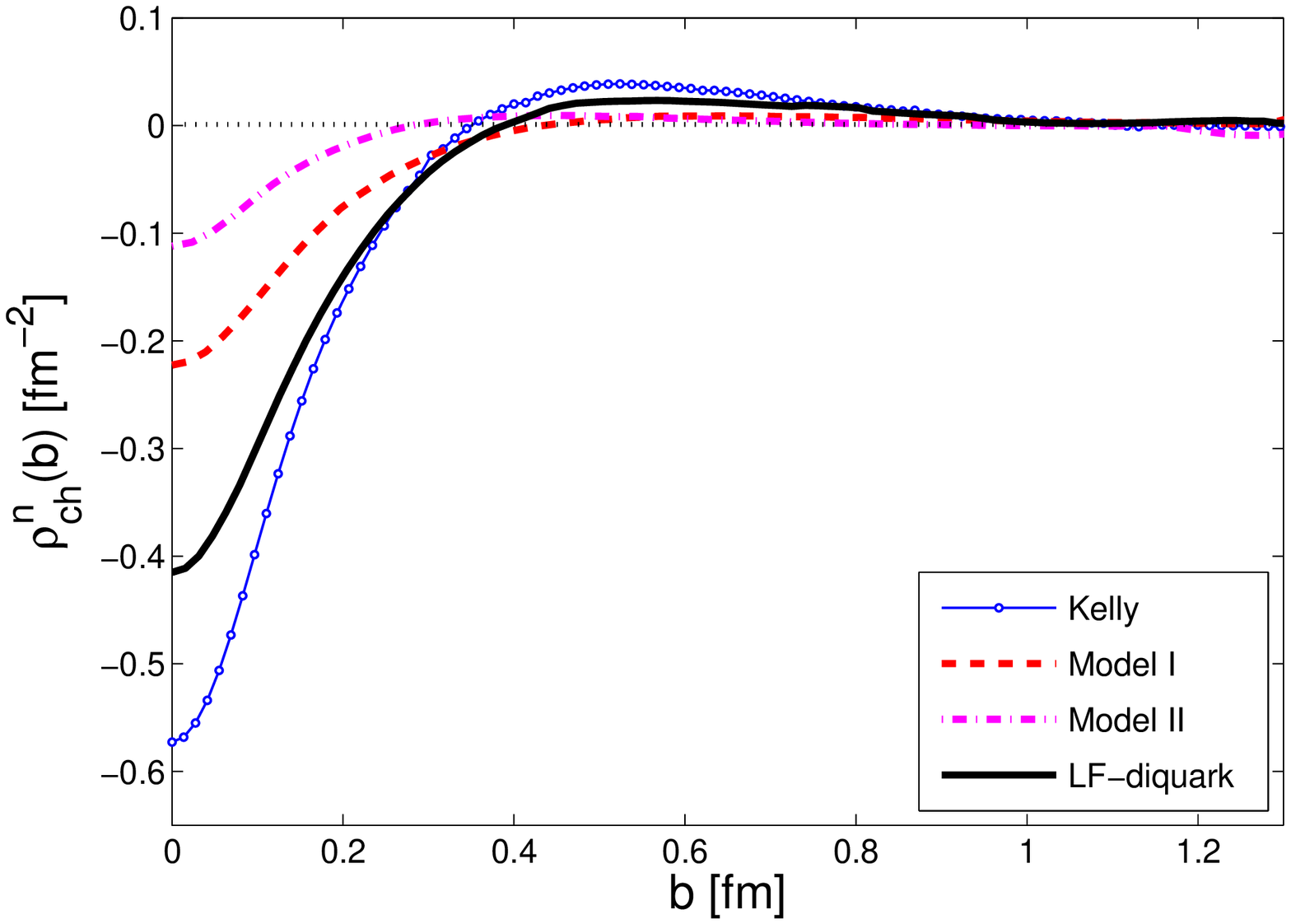}
\hspace{0.1cm}%
\small{(d)}\includegraphics[width=5cm,clip]{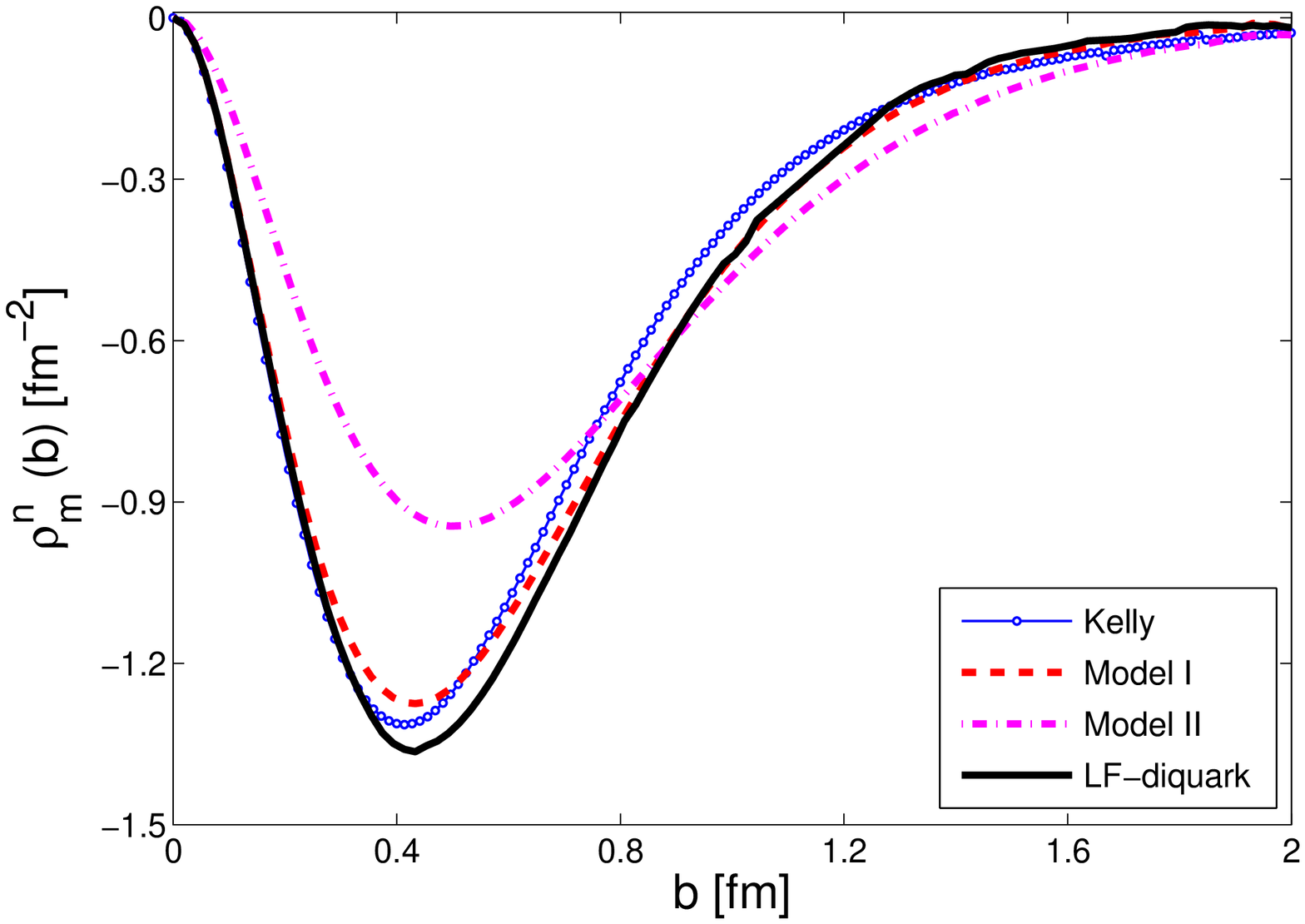}
\end{minipage}
\caption{\label{nucleon_den} Transverse charge and anomalous magnetization densities for nucleon. (a) and (b) represent $\rho_{ch}$ and ${\rho}_m$ for the proton. (c) and (d) same as proton but for neutron. Lines with circle represent the parametrization of Kelly \cite{kelly04}.}
\end{figure*} 
\begin{figure*}[htbp]
\begin{minipage}[c]{0.98\textwidth}
\small{(a)}
\includegraphics[width=5cm,clip]{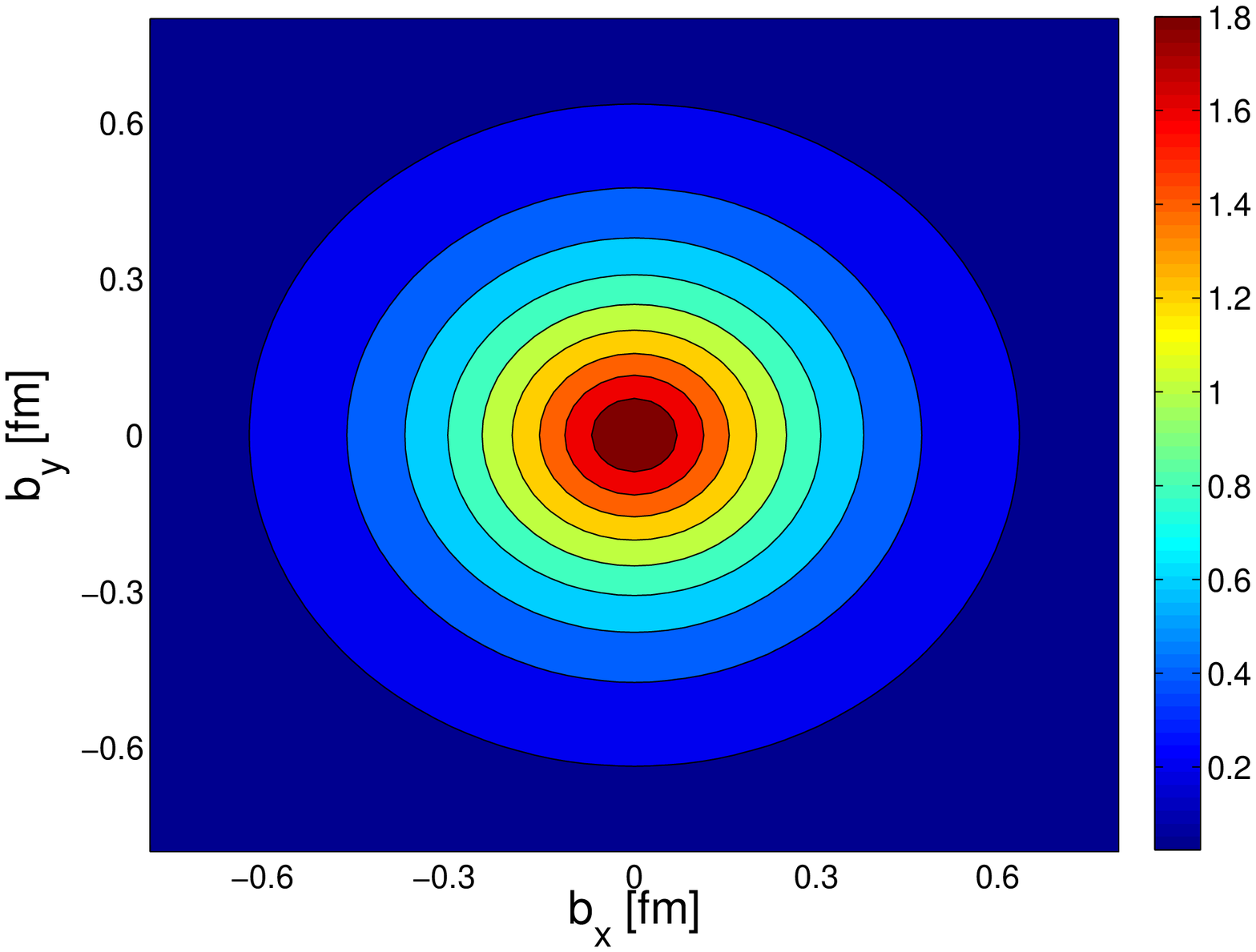}
\hspace{0.1cm}%
\small{(b)}\includegraphics[width=5cm,clip]{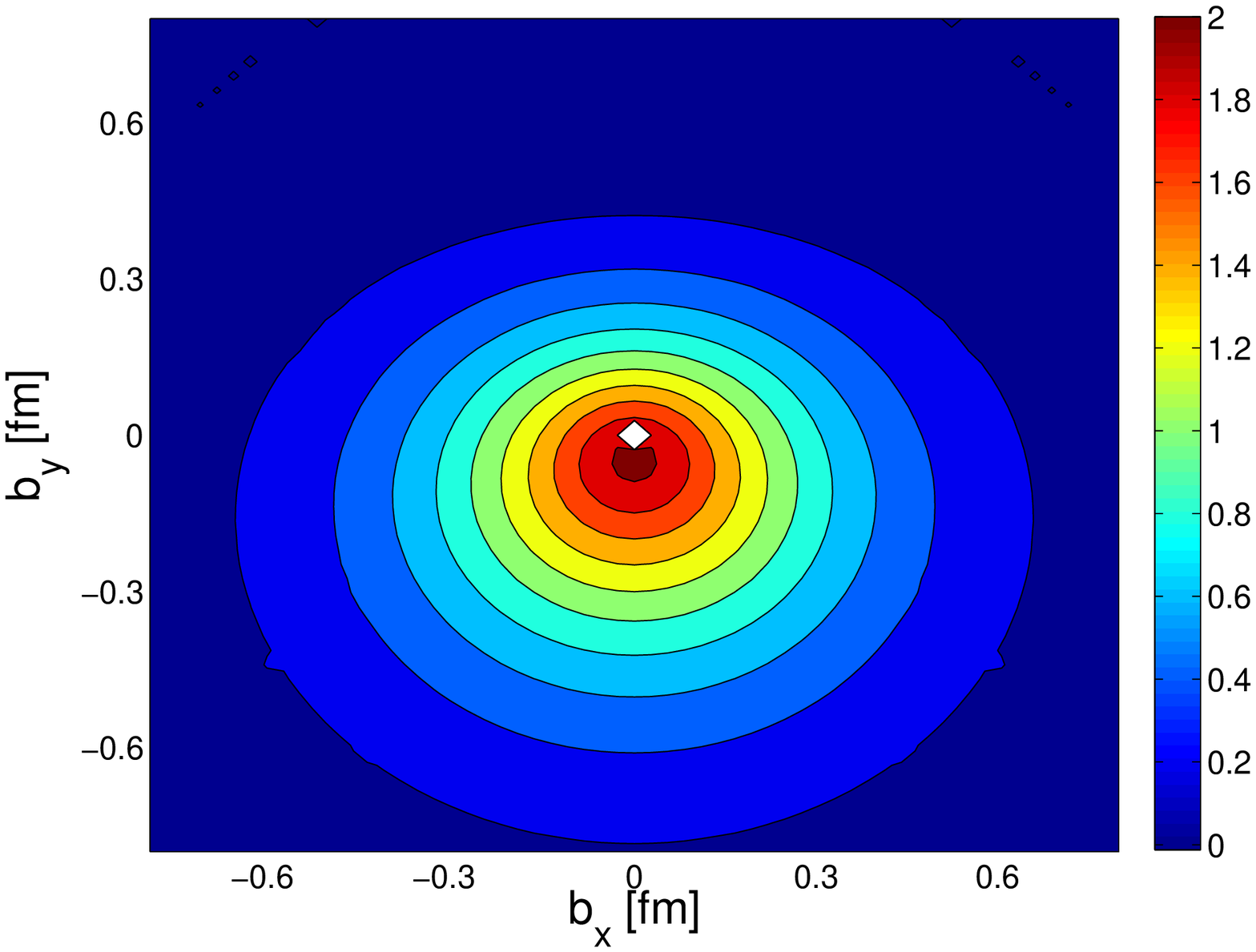}
\end{minipage}
\begin{minipage}[c]{0.98\textwidth}
\small{(c)}
\includegraphics[width=5cm,clip]{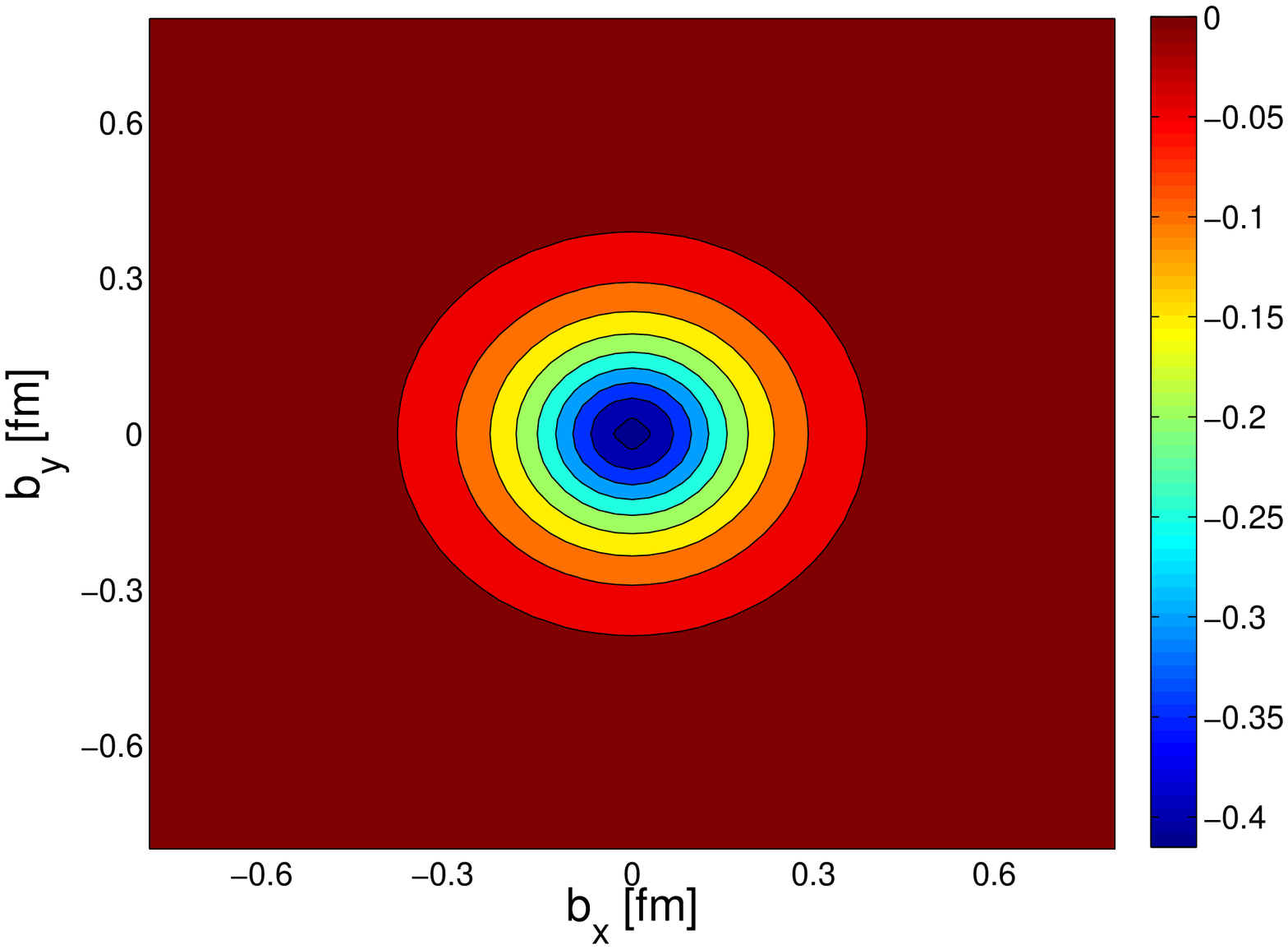}
\hspace{0.1cm}%
\small{(d)}\includegraphics[width=5cm,clip]{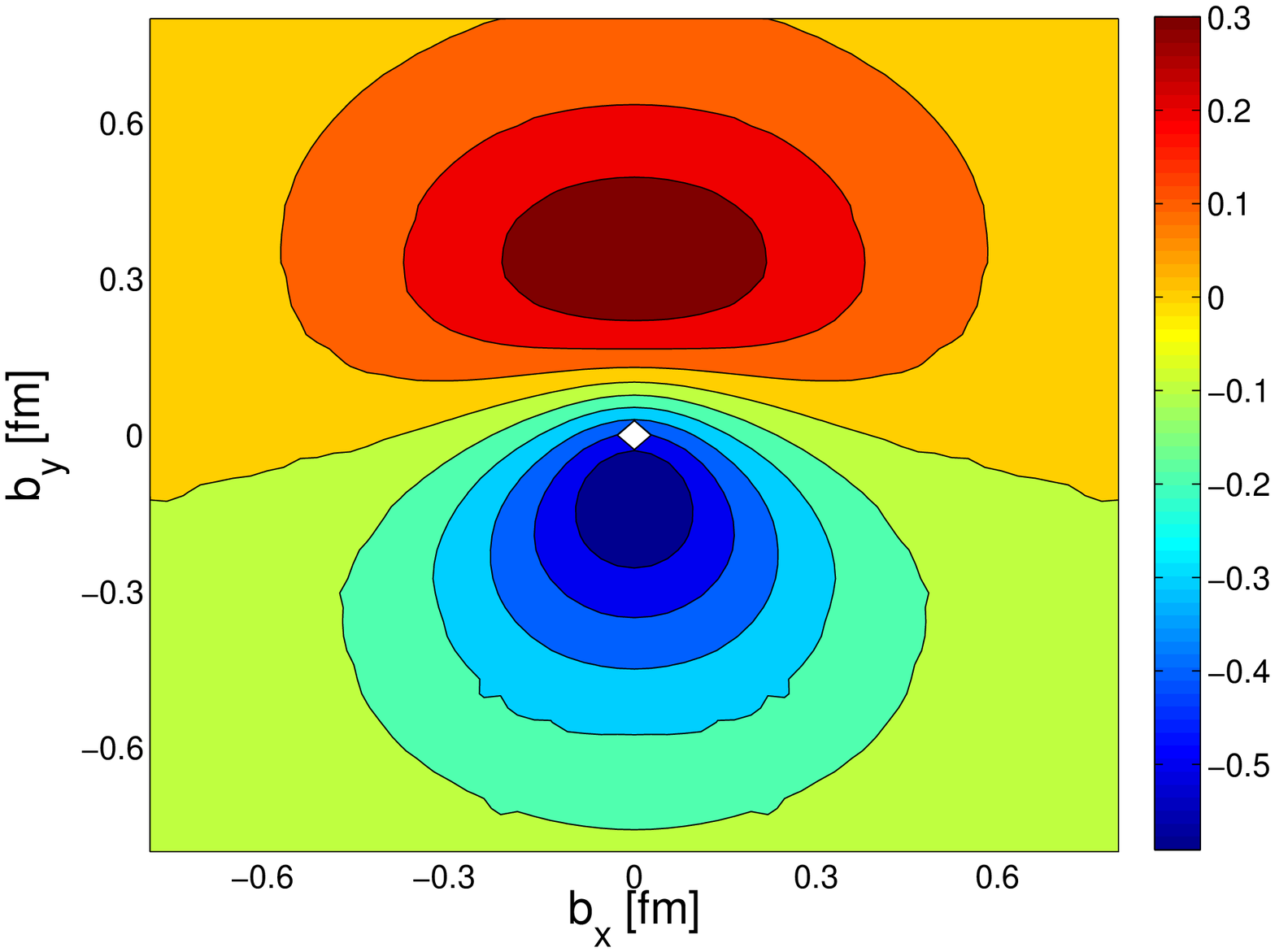}
\end{minipage}
\caption{\label{nucleon_den_3D}Unpolarized and transversely polarized charge densities for proton(upper) and neutron(lower) in LF diquark model.}
\end{figure*} 

{\bf Model II} : 
The other model of the nucleon form factors was formulated by Abidin and Carlson\cite{AC}. Since the action defined in Eq.(\ref{action}) can not generate the Pauli form factors, they introduced an additional gauge invariant non-minimal coupling term 
\be\label{F2AdS}
\int d^4x~dz~\sqrt{g}  ~ \bar\Psi
 \,  e_M^A\,  e_N^B \left[\Gamma_A, \Gamma_B\right] F^{M N}\Psi .
 \ee
 This additional term also provides an anomalous contribution to the Dirac form factor. In this model the form factors are given by\cite{AC}
 \be
 F_1^p(Q^2) &=& C_1(Q^2)+\eta_p C_2(Q^2),\label{F1pM2}\quad \quad F_1^n(Q^2)= \eta_n C_2(Q^2),\label{F1nM2}\\
 F_2^p(Q^2)&=& \eta_p C_3(Q^2),\label{F2pM2}\quad \quad \quad \quad \quad \quad ~ F_2^n(Q^2)= \eta_n C_3(Q^2).\label{F2nM2}
 \ee
  The functions $C_i(Q^2)$ are defined as 
 $C_1(Q^2) = \frac{a+6}{(a+1)(a+2)(a+3)},$
 $C_2(Q^2) =\! \frac{2a(2a-1)}{(a+1)(a+2)(a+3)(a+4)},$ and
$ C_3(Q^2) =\! \frac{48}{(a+1)(a+2)(a+3)}, $
 where $a=Q^2/(4\kappa^2)$. The value of  $\kappa=0.350 GeV$  is fixed by simultaneous fit to proton and rho meson masses. The other parameters are determined from the normalization conditions of the Pauli form factor at $Q^2=0$ and are given by $\eta_p=0.224$ and $\eta_n=-0.239$ \cite{AC}. 
The Pauli form factors in Model I and Model II are identical, the main difference is in the Dirac form factor where we have 
 an additional contribution in Model II coming the extra term added.

{\bf Quark-diquark model in AdS/QCD :}
Here we consider a light front  scalar quark-diquark model for nucleon\cite{Gut} where the 2-particle wavefunction is modeled from the soft-wall AdS/QCD solution. In the light front overlap formalism, the electromagnetic form factors in this model are given by
\be
F_1^q(Q^2)&=& \int_0^1dx \int \frac{d^2\bfk}{16\pi^3}~\bigg[\psi_{+q}^{+*}(x,\bfk')\psi_{+q}^+(x,\bfk)
+\psi_{-q}^{+*}(x,\bfk')\psi_{-q}^+(x,\bfk)\bigg],\\
F_2^q(Q^2)&=& -\frac{2M_n}{q^1-iq^2}\int_0^1dx \int \frac{d^2\bfk}{16\pi^3}~\bigg[\psi_{+q}^{+*}(x,\bfk')\psi_{+q}^-(x,\bfk)+\psi_{-q}^{+*}(x,\bfk')\psi_{-q}^-(x,\bfk)\bigg],
\ee
where $\bfk'=\bfk+(1-x)\bfq$. $\psi_{\lambda_q q}^{\lambda_N}(x,\bfk)$ are the LFWFs with specific nucleon helicities $\lambda_N=\pm$ and for the struck quark $\lambda_q=\pm$, where plus and minus correspond to $+\frac{1}{2}$ and $-\frac{1}{2}$ respectively. The LFWFs are 
specified at an initial scale $\mu_0=313$~MeV \cite{Gut} :  
\be 
\psi_{+q}^+(x,\bfk) &=&  \varphi_q^{(1)}(x,\bfk) 
\,, \quad \quad \quad \quad \quad \quad
\psi_{-q}^+(x,\bfk) = -\frac{k^1 + ik^2}{xM_n}   \, \varphi_q^{(2)}(x,\bfk) \,, \nonumber\\
\psi_{+q}^-(x,\bfk) &=& \frac{k^1 - ik^2}{xM_n}  \, \varphi_q^{(2)}(x,\bfk)
\,, \quad \quad
\psi_{-q}^-(x,\bfk) = \varphi_q^{(1)}(x,\bfk) 
\,, \label{WF} \\
\varphi_q^{(i)}(x,\bfk)&=&N_q^{(i)}\frac{4\pi}{\kappa}\sqrt{\frac{\log(1/x)}{1-x}}x^{a_q^{(i)}}(1-x)^{b_q^{(i)}}\exp\bigg[-\frac{\bfk^2}{2\kappa^2}\frac{\log(1/x)}{(1-x)^2}\bigg].
\ee
For $a_q^{(i)}=b_q^{(i)}=0$, $\varphi_q^{(i)}(x,\bfk)$ reduces to the AdS/QCD prediction \cite{BT2}. $\kappa$ is the AdS/QCD scale parameter which is taken to be $0.4$ GeV \cite{CM1,CM2}. 
 The   parameters $a^{(i)}_q$ and $b^{(i)}_q$ with the constants $N^{(i)}_q$  are fixed by fitting the electromagnetic properties of the nucleons \cite{CM6}. 
    \begin{table}[t]
\caption{Electromagnetic radii of the nucleons}
\centering
\label{tab:1}       
\begin{tabular}{lllll}
\hline\noalign{\smallskip}
Quantity & Model I & Model II & LF diquark & Measured data \\ [5pt] 
\tableheadseprule\noalign{\smallskip}
$r_E^p$ (fm) & 0.810 & 0.980 & 0.786   & $0.877 \pm 0.005$ \\ 

$r_M^p$ (fm) & 0.782 & 0.921 & 0.772 & $0.777\pm0.016$  \\ 
\hline 
$\langle r_E^2\rangle^n$ (fm$^2$) & $-0.088$ & $-0.123$ & $-0.085 $ & $-0.1161 \pm 0.0022$\\

$r_M^n$ (fm) & 0.796 & 0.937 & 0.7596 & $0.862^{+0.009}_{-0.008} $\\
\noalign{\smallskip}\hline
\end{tabular}
\end{table}

{\bf Flavor decompositions of the nucleon form factors :}
 Under  the charge  and isospin symmetry it is straightforward to write down the flavor decompositions of the nucleon form factors as \cite{Cates}
\be
F_i^p=e_u F_i^u + e_d F_i^d ~~{\rm and} ~~F_i^n=e_d F_i^u + e_u F_i^d,~~(i=1,2)
\ee
where $e_u$ and $e_d$ are charge of $u$ and $d$ quarks respectively. The normalizations of the flavor form factors are $F_1^u(0)=2, F_2^u(0)=\kappa_u$ and $F_1^d(0)=1, F_2^d(0)=\kappa_d$ where the anomalous magnetic moments for the up and down quarks are $\kappa_u=2\kappa_p+\kappa_n=1.673$ and $\kappa_d=\kappa_p+2\kappa_n=-2.033$. 

In Fig.\ref{proton_fit} we compare the results for electromagnetic form factors of nucleons calculated in different AdS/QCD models. The flavor form factors are shown in Fig.\ref{FF_flavors}. The figures show that the results of the Model I and the quark-diquark model are in good agreement with experimental data whereas the Model II deviates from the data. Only for $F_1^d$, Model I deviates at higher $Q^2$ from the data and also for $G_E^n(Q^2)$ Model II is better than Model I. The fitted results for the electromagnetic radii of the nucleons are listed in Table \ref{tab:1}. 
\section{Transverse charge and magnetization densities} \label{sec:2}
The transverse charge density inside the nucleons is given by 
\be
\rho_{ch}(b)
=\int \frac{d^2q_{\perp}}{(2\pi)^2}F_1(q^2)e^{iq_{\perp}.b_{\perp}}
=\int_0^\infty \frac{dQ}{2\pi}QJ_0(Qb)F_1(Q^2),
\ee
where $b$ represents the impact parameter and $J_0$ is the cylindrical Bessel function of order zero. 
One can define the magnetization density($\widetilde{\rho}_{M}(b)$) in the similar fashion with $F_1$ is replaced
by $F_2$,
whereas,
$\rho_m(b)= -b(\partial \widetilde{\rho}_M(b)/\partial b)$
can be interpreted as anomalous magnetization density. 
We  evaluate the quark contributions to the nucleon transverse densities using charge and isospin symmetry  (see \cite{CM3,chandan}).

For transversely polarized nucleon, the charge density is given by 
\be
\rho_T(b)=\rho_{ch}-\sin(\phi_b-\phi_s)\frac{1}{2M_n b}\rho_m\label{trans_pol}.
\ee
The transverse polarization of the nucleon is given by
$S_\perp=(\cos\phi_s \hat{x}+\sin\phi_s\hat{y})$ and the transverse impact parameter $b_\perp=b(\cos\phi_b \hat{x} +\sin\phi_b\hat{y})$. Without loss of generality, the polarization of the nucleon is taken along $x$-axis ie., $\phi_s=0$. The second term in Eq.(\ref{trans_pol}), provides the deviation from circular symmetry of the unpolarized charge density.
   The nucleon charge and anomalous magnetization densities presented in Fig.\ref{nucleon_den}  suggest that the quark-diquark model agrees with the phenomenological parametrizations \cite{kelly04} much better than the Model I and the Model II and Model I is better compare to Model II. 
 The charge densities between unpolarized and transversely polarized nucleon in the quark-diquark model are compared in Fig.\ref{nucleon_den_3D}. The unpolarized densities are axially symmetric in transverse plane while for the transversely polarized nucleons they become distorted. For nucleon polarized along $x$ direction, the densities get shifted towards negative $y$-direction. Due to large anomalous magnetization density, the distortion in neutron charge density is found to be stronger than that for proton.

\begin{acknowledgements}
CM thanks the Sciencce and Engineering Research Board (SERB), Gorernment of India for supporting the travel grant under contract no. ITS/3444/2015-2016 to attend the conference LightCone2015  where the work was presented.
\end{acknowledgements}



\end{document}